\documentclass[
nobibnotes,aps,pre,showkeys,
12pt]{revtex4-2}
\usepackage{hyperref}
\usepackage{graphics}
\usepackage{subcaption}
\usepackage{graphicx}
\usepackage{color}
\usepackage{amsmath}
\usepackage{amssymb}
\usepackage{bm}
\usepackage{float}
\usepackage{epsfig}
\usepackage{epstopdf}
\usepackage{color,graphics}
\usepackage{mathtools}

\begin{document}

\title{
Noether symmetries and conservation laws of a reduced gauged bilayer graphene model}
\author{Fernando Haas}
\affiliation{Physics Institute, Federal University of Rio Grande do Sul, 91501-970,
Av. Bento Gon\c{c}alves 9500, Porto Alegre, RS, Brazil}
\email{fernando.haas@ufrgs.br}


\begin{abstract}
A canonical Hamiltonian is found for a reduced version of the Jackiw-Pi model for bilayer graphene. From the corresponding Lagrangian, the Noether point symmetries and conserved quantities are determined. The Noether symmetry group is the same as the fifth-dimensional group for the time-dependent harmonic oscillator. The realization of the algebra is achieved in terms of just one particular solution $g_1$ of the time-dependent harmonic oscillator equation underlying the reduced Jackiw-Pi model. Some numerical solutions are worked out. 
\end{abstract}


\maketitle

\section{Introduction}

The experimental realization of stable graphene \cite{Novo} is a current research topic specially in view of 
possible new technologies such as new electronic devices. The concrete realization of Dirac zero modes and fractional charge \cite{Dirac} can be also pursued, with the low-energy spectrum of graphene described by the Dirac equation in two spatial dimensions. The existence of zero-energy modes in monolayer graphene in principle can be obtained if the Dirac Hamiltonian also has an interaction with a scalar potential, in terms of the Kekul\'e dimerization \cite{five}. 
This dimerization can be difficult to be produced, inspiring the investigation of bilayer graphene where the two layers are superposed and separated by a dielectric barrier. A gauged version of this system  has an order parameter playing the same r\^ole of the Kekul\'e distortion \cite{Jackiw}, allowing for topological effects, fractional charge and zero modes. 
The resulting Jackiw-Pi model is modeled by a Dirac equation in two spatial dimensions where the four-spinor describes the two Dirac points and the two sublattices (Eq. (2) of \cite{Jackiw}). Separating the angular dependence one has a reduced version which is the radial Dirac equation in this case \cite{Jackiw, Rosu}. The purpose of the present work is to perform the Noether symmetry analysis of this reduced Jackiw-Pi model, thanks to the identification of a suitable canonical action. 
Symmetry is recognized as a fundamental subject, with Noether theorem providing a link between the existence of conserved quantities and symmetries \cite{Noet}. Here the treatment will be restricted to Noether point (or geometrical) symmetries. 

In \cite{saidecima} symmetries of the classical action for the Jackiw-Pi in 2 + 1 space-time dimensions  have been enumerated, without direct reference to Noether's theorem. However, there is no systematic search for Noether point symmetries as we did, for the reduced Jackiw-Pi model presented below in Eqs. (\ref{e1}) and (\ref{e2}). 

This work is organized as follows. In Section II, a canonical action in configuration space is identified for the reduced Jackiw-Pi model. This allows the search of Noether point symmetries and invariants, which is shown to be determined by the solution of a third-order and a second-order linear ordinary differential equations. The corresponding conserved quantities are obtained and interpreted. In Section III the concrete realization of the symmetry algebra is shown to be achieved in terms of a single particular solution of the time-dependent harmonic oscillator (TDHO) equation which is the same equation underlying the reduced Jackiw-Pi model. Section IV presents the direct approach towards the derivation of the full Noether symmetries and conserved quantities, after the choice of the appropriate gauge field entering the model. The singular point of the basic TDHO equation at zero time is identified. Section V shows numerical examples and Section VI contains the conclusion and final remarks.

\section{Noether symmetries. Conservation laws}

A reduced version of the Jackiw-Pi model for bilayer graphene \cite{Jackiw} is given by 
\begin{eqnarray}
\label{e1}
\left(D_r - \Phi(r)\right)u(r) &=& V\,v(r) \,,\\
\label{e2}
\left(D_r + \Phi(r)\right)v(r) &=& - V\,u(r) \,,
\end{eqnarray}
where $D_r = d/dr$ and
\begin{equation}
\label{gau}
\Phi(r) = \frac{n}{r}\left(\frac{1}{2} - A(r)\right) \,,  
\end{equation}
where  $A(r)$ is a gauge field such that $A(0) = 0, A(\infty) = 1/2$ and $V$ is an applied bias potential, here assumed to be positive for the sake of definition, while $n$ appears in the angular part of the scalar field related to the interactions between electrons in one layer and holes in the other. 
The reasoning towards Eqs. (\ref{e1})-(\ref{e2}) starts from the Dirac equation in two spatial dimensions, 
describing low-energy excitations of bilayer graphene. Assuming a special form of the Dirac spinor with a vortex-like angular dependence (Eqs. (18a)-(18b)   of Ref. \cite{Jackiw}) the result is the system (\ref{e1})-(\ref{e2}), where $r$ is the radial coordinate. Single-valuedness of the Dirac spinor requires $n$ to be an odd integer. So instead of the full Dirac equation one has a reduced model which is a pair of first-order ordinary differential equations for $u, v$. As shown in \cite{Rosu}, $u$ and $v$ can be viewed as supersymmetric partners in supersymmetric quantum mechanics. 



We set
\begin{equation}
\label{phi}
q = u \,, \quad p = v \,, \quad t = V r \,, \quad \phi(t) = \frac{\Phi(r)}{V} 
\end{equation}
to obtain 
\begin{equation}
\label{e3}
\dot{q} = p + \phi(t)\,q \,, \quad \dot{p} = - q - \phi(t)\,p \,,
\end{equation}
where the dot denotes derivative with respect to the independent variable $t \geq 0$, which formally plays the role of time. 

Notice that $q, p$ satisfy two uncoupled TDHO equations \cite{Rosu}, namely
\begin{equation}
\label{qp}
\ddot{q} + (1 - \dot\phi - \phi^2)\,q = 0 \,, \quad \ddot{p} + (1 + \dot\phi - \phi^2)\,p = 0 \,. 
\end{equation}
As apparent, the equation for the partner variable $p$ is the same as the equation for $q$, provided the sign of $\phi = \phi(t)$ is reversed. 

It happens that (\ref{e3}) has a Hamiltonian form, 
\begin{equation}
\dot{q} = \frac{\partial H}{\partial p} \,, \quad \dot{p} = - \frac{\partial H}{\partial q} \,,
\end{equation}
with the quadratic time-dependent Hamiltonian function
\begin{equation}
H = \frac{1}{2}(p^2+q^2) + \phi(t)\,p\,q \,.
\end{equation}
This suggest a search for constants of motion using Noether's theorem, which apply in a more direct way using the associated Lagrangian $L = p\,\dot{q} - H$, viz. 
\begin{equation}
\label{l}
L = \frac{\dot{q}^2}{2} - \frac{1}{2}(1-\phi^{2}(t))\,q^2 - \phi(t)\,q\,\dot{q} \,.
\end{equation}

Noether's theorem \cite{Noet, Cantrijn} states that if the action functional $S = \int_{t1}^{t2} L\, dt$ remains invariant (up to addition of a numerical constant) under an  infinitesimal point transformation
\begin{equation}
t \rightarrow t + \varepsilon\,\tau(q,t) \,, \quad q \rightarrow q + \varepsilon\,\eta(q,t) \,,
\end{equation}
where $0 < \varepsilon \ll 1$, then the quantity 
\begin{equation}
\label{noe}
I = \tau\left(\dot{q}\,\frac{\partial L}{\partial\dot{q}} - L\right) - \eta\,\frac{\partial L}{\partial\dot{q}} + F \,.
\end{equation}
is a constant of motion ($dI/dt = 0$), the Noether invariant. The condition of (quasi) invariance of the action is 
\begin{equation}
\label{inv}
\tau\frac{\partial L}{\partial t} + \eta\frac{\partial L}{\partial q} + (\dot\eta - \dot\tau\dot{q})\frac{\partial L}{\partial\dot{q}} + \dot{\tau}L = \dot{F} \,,
\end{equation}
where $F = F(q,t)$ is a function to be determined. In Eq. (\ref{inv}) one has e.g. $\dot{F} = F_t + F_q\,\dot{q}$, where at this stage it is useful to denote partial derivatives by a subscript. For instance, $F_t = \partial F/\partial t, F_q = \partial F/\partial q$.

For later reference, the generator of Noether point symmetries is 
\begin{equation}
\label{gg}
G = \tau(q,t)\,\frac{\partial}{\partial t} + \eta(q,t)\,\frac{\partial}{\partial q} \,.
\end{equation}

The procedure for calculating Noether point symmetries is well known \cite{Cantrijn}. 
In the present case inserting the Lagrangian (\ref{l}) into the invariance condition (\ref{inv}) gives that a third-order polynomial in $\dot{q}$ must vanishes identically. The vanishing of each monomial 
gives: 
\begin{subequations}
\begin{align}
\label{a}
\dot{q}^3   & \Rightarrow \quad \tau_q = 0 \,,    \\
\label{b}
\dot{q}^2  &  \Rightarrow \quad \eta_q - \tau_t/2 = 0 \,,  \\
\label{c}
\dot{q}  &  \Rightarrow \quad -\tau\,\dot\phi\,q - \eta\,\phi + \eta_t - \phi\,q\,\eta_q = F_q \,, \\
\label{d}
\dot{q}^{(0)} & \Rightarrow \quad \tau\,\phi\,\dot\phi\,q^2 - \eta\,(1-\phi^2)\,q - \phi\,q\,\eta_t - (1 - \phi^2)\,q^2\,\tau_t/2  = F_t \,. 
\end{align}
\end{subequations}

From Eqs. (\ref{a}) and (\ref{b}) it follows that  
\begin{equation}
\tau = T(t) \,, \quad \eta = \dot{T}(t)\,q/2 - g(t) \,,
\end{equation}
where $T = T(t)$ and $g = g(t)$ are functions depending on time only. Then from Eqs. (\ref{c}) and (\ref{d}) together with $F_{qt} = F_{tq}$ yields the vanishing of a linear polynomial in $q$. The vanishing of the corresponding monomials yields
\begin{subequations}
\begin{align}
\label{aa}
q   & \Rightarrow \quad 
\dddot T + 4\,\omega^2\,\dot T + 4\,\,\omega\,\dot\omega\,T = 0  \,,    \\
\label{bb}
q^{(0)}  &  \Rightarrow \quad \ddot{g} + \omega^{2}\,g = 0 \,,  
\end{align}
\end{subequations}
where the time-dependent oscillator frequency $\omega = \omega(t)$ comes from 
\begin{equation}
\label{ome}
\omega^{2} = 1 -\dot\phi - \phi^2 \,.
\end{equation}
Incidentally, the auxiliary variable $g$ satisfies the same equation of the dynamical variable $q$, as seen from Eqs. (\ref{qp}) and (\ref{bb}).

Setting $T = \rho^2$ yields 
\begin{equation}
\rho\,\dddot\rho + 3\,\dot\rho\,\ddot\rho + 4\,\omega^2\,\rho\,\dot\rho + 2\,\omega\,\dot\omega\,\rho^2 = 0 \,,
\end{equation}
which can be integrated once to give Pinney's \cite{Pinney}   equation 
\begin{equation}
\label{pin}
\ddot\rho +  \omega^{2}\,\rho = \kappa/\rho^3 \,,
\end{equation}
where $\kappa$ is an arbitrary numerical constant. 

From Eqs. (\ref{c}) and (\ref{d}) one then finds 
\begin{equation}
F = \left(\frac{\ddot T}{2} - \phi\,\dot T - \dot\phi\,T\right)\frac{q^2}{2} - (\dot g - \phi\,g)\,q + F_0 \,,
\end{equation}
where $F_0$ is a numerical constants which will be set to zero without loss of generality. 


Correspondingly, using Eq. (\ref{noe}) the Noether invariant is given by
\begin{equation}
\label{n}
I = \frac{T}{2}\,{\dot q}^2 - \frac{\dot T}{2}\,q\,\dot q + \left(\ddot T + 2\,\omega^2\,T\right)\frac{q^2}{4} +  g\,\dot{q} - \dot{g}\,q \,.
\end{equation}
%

%
%

From Eq. (\ref{gg}) the generator of Noether point symmetries becomes 
\begin{equation}
\label{gt}
G = T\,\frac{\partial}{\partial t} + \left(\frac{\dot T}{2}\,q - g\right)\,\frac{\partial}{\partial q} \,.
\end{equation}
From the total of five linearly independent solutions of the linear equations (\ref{aa}) and (\ref{bb}), one naturally has a 5-parameter group of Noether point symmetries, whose five generators are readily found from Eq. (\ref{gt}).  To each generator there is one conserved quantity from Eq. (\ref{n}). Among these five invariants only two are functionally independent. Notice that the asymptotes of the gauge field $A(r)$ in Eq. (\ref{gau}) limits the admissible Noether symmetries and invariants.  


In terms of the solution $\rho$ of the Pinney equation, so that $T = \rho^2$, one has 
\begin{equation}
\label{nn}
I = I_{\rm EL} + W \,, 
\end{equation}
where 
\begin{equation}
\label{EL}
I_{\rm EL} = \frac{1}{2}(\rho\,\dot{q} - \dot\rho\,q)^2 + \frac{\kappa}{2}\,\left(\frac{q}{\rho}\right)^2 \,, \quad W = g\,\dot{q} - \dot{g}\,q 
\end{equation}
are two functionally independent invariants. 
The conserved quantity (\ref{nn}) consists of two parts, namely the Ermakov-Lewis part $I_{\rm EL}$ and the remaining contribution $W$ which is non-zero for nontrivial $g$. The later part arises because both $q$ and $g$ satisfy the same TDHO equation, as seen from Eqs. (\ref{qp}) and (\ref{bb}), and the Wronskian of two independent particular solutions of this linear second-order equation is a constant of motion. As remarked e.g. in \cite{Leone}, the constancy of the Wronskian in this case is not an accidental fact, but a result from a symmetry invariance. 

In this version, 
\begin{equation}
G = \rho^2\,\frac{\partial}{\partial t} + (\rho\dot\rho\,q - g)\,\frac{\partial}{\partial q} \,.
\end{equation}
%
However, the less traditional expressions (\ref{n}) and (\ref{gt}) show in manifest way how the search for Noether invariants and symmetries is reducible in this case to a pair of linear equations, the third-order equation (\ref{aa})  for $T$ and the second-order equation (\ref{bb}) for $g$. 



\section{Realization of the symmetry algebra and Noether invariants}

Suppose one knows a exact particular nontrivial solution $g_1 = g_1(t)$ for the TDHO equation (\ref{bb}), which is formally the same equation satisfied by the dynamical variable $q(t)$. As is well-known, a second, linearly independent solution of the TDHO is then given by 
\begin{equation}
\label{g2}
g_2 = g_2(t) = g_1(t)\,\int\,dt/g_{1}^{2}(t) \,.
\end{equation}
This pair of solutions have a unit Wronskian $g_1\dot g_2 - g_2\dot g_1 = 1$. The general solution of Eq. (\ref{bb}) is 
\begin{equation}
g = c_1\,g_1 + c_2\,g_2 \,,
\end{equation}
where $c_{1,2}$ are arbitrary numerical constants. Of course a different pair of fundamental solutions could be chosen, but having a unit Wronskian simplify several numerical factors in the continuation. Given $g_1$, Eq. (\ref{g2}) for $g_2$ will be assumed henceforth. 

The general solution of Pinney's equation (\ref{pin}) is given \cite{Pinney} by the nonlinear superposition law 
\begin{equation}
\label{nlsl}
\rho = (c_3\,g_1^2 + 2\,c_4\,g_1\,g_2 + c_5\,g_2^2)^{1/2} \,,
\end{equation}
where $c_{3,4,5}$ are numerical constants satisfying $c_3\,c_5 - c_4^2 = \kappa$.

From $T = \rho^2$ one then derive the general solution of the linear equation (\ref{aa}) in the form 
\begin{equation}
\label{tt}
T = c_3\,T_1 + c_4\,T_2 + c_5\,T_3 \,,
\end{equation}
where 
\begin{equation}
T_1 = g_1^2 \,, \quad T_2 = 2\,g_1\,g_2 \,, \quad T_3 = g_2^2 
\end{equation}
are linearly independent particular solutions. Inserting $T_{1,2,3}$ indeed solves Eq. (\ref{aa}), provided $g_1$ solves the TDHO equation (\ref{bb}) and $g_2$ is given by Eq. (\ref{g2}).  

Instead of the traditional approach starting from the {\it nonlinear} Pinney equation for a given $\kappa$, we focus on the {\it linear} third-order equation (\ref{aa}), which we show 
to be exactly solvable provided {\it only one} particular solution of the TDHO equation is available.  The parameter $\kappa$ is determined from $c_{3,4,5}$ arbitrarily chosen. 

From this reasoning, the symmetry generator in Eq. (\ref{gt}) is
\begin{equation}
G = c_1\,G_1 + c_2\,G_2 + c_3\,G_3  + c_4\,G_4 + c_5\,G_5  \,,
\end{equation}
where
\begin{subequations}
\begin{align}
\label{g1}
G_1 & = - g_1\,\frac{\partial}{\partial\,q} \,,    \\
G_2 & = - g_2\,\frac{\partial}{\partial\,q} \,, \\
G_3 & =  g_1^2\,\frac{\partial}{\partial\,t} + g_1\,\dot g_1\,q\,\frac{\partial}{\partial\,q}  \,, \\
G_4 & = 2\,g_1\,g_2\,\frac{\partial}{\partial\,t} + (g_1\,\dot g_2 + g_2\,\dot g_1)\,q\,\frac{\partial}{\partial\,q} \,, \\
\label{g5} 
G_5 & =  g_2^2\,\frac{\partial}{\partial\,t} + g_2\,\dot g_2\,q\,\frac{\partial}{\partial\,q}  \,. 
\end{align}
\end{subequations}

The corresponding Noether invariant in Eq. (\ref{n}) is 
\begin{equation}
I = c_1\,I_1 + c_2\,I_2 + c_3\,I_3  + c_4\,I_4 + c_5\,I_5  \,,
\end{equation}
where each $I_i$ is associated to $G_i, i = 1,\dots,5$ and 
\begin{subequations}
\begin{align}
\label{i1}
I_1 & = g_1\,\dot q - \dot g_1\,q \,,    \\
I_2 & = g_2\,\dot q - \dot g_2\,q  \,, \\
I_3 & = \frac{1}{2}\,(g_1\,\dot q - \dot g_1\,q)^2 = \frac{I_1^2}{2} \,, \\
I_4 & = g_1\,g_2\,\dot{q}^2 - (g_1\,\dot g_2 + g_2\,\dot g_1)\,q\,\dot q + \dot{g}_1\,\dot{g}_2\,q^2 = I_1\,I_2 \,, \\
\label{i5} 
I_5 & = \frac{1}{2}\,(g_2\,\dot q - \dot g_2\,q)^2 = \frac{I_2^2}{2} \,.
\end{align}
\end{subequations}

The Ermakov-Lewis invariant from Eq. (\ref{EL}) with $\rho$ given by Eq. (\ref{nlsl}) can be expressed as 
\begin{equation}
I_{\rm EL} = c_3\,I_3 + c_4\,I_4 + c_5\,I_5 \,,
\end{equation}
which could be expected a priori and can be also verified by direct calculation to be a constant of motion.  In this manner it is seen that the Ermakov-Lewis invariant can be split into a linear combination of quadratic functions of the conserved Wronskians. 
The remaining Wronskian part of the Noether invariant is 
\begin{equation}
W = c_1\,I_1 + c_2\,I_2 \,.
\end{equation}
%


From the commutator $[G_i, G_j] = G_i\,G_j - G_j\,G_i$ one finds the symmetry algebra 
\begin{eqnarray}
&\strut& [G_1, G_2] = [G_1, G_3] = [G_2, G_4] = [G_2, G_5] = 0 \,, \quad [G_1, G_4] = G_1 \,, \quad [G_1, G_5] = G_2 \,, \nonumber \\ 
&\strut& [G_2, G_3] = - G_1 \,, \quad  [G_3, G_4] = 2\,G_3 \,, \quad  [G_3, G_5] = G_4 \,, \quad  [G_4, G_5] = 2\,G_5 \,. 
\label{sa}
\end{eqnarray}

The simplest description of the Noether symmetries and invariants if provided by the invariants $I_{1,2}$ with corresponding commuting symmetry generators $G_{1,2}$ expanding an Abelian two-dimensional subgroup. The remaining $I_{1,2,3}$ are quadratic functions of the building blocks $I_{1,2}$. 

The fifth-dimensional group of Noether points symmetries is a subgroup of the maximal eight-dimensional Lie group ${\rm SL}(3,\Re)$ admitted by the simple harmonic oscillator \cite{Wybourne, Lie} and the TDHO \cite{Leach}. The connection with the TDHO  comes because the Lagrangian in Eq. (\ref{l}) is 
\begin{equation}
L = {\cal L} - \frac{d}{dt}\left(\frac{\phi(t)\, q^2}{2}\right)
\end{equation}
where 
\begin{equation}
\label{stla}
{\cal L} = \frac{1}{2}\,{\dot q}^2 - \frac{1}{2}\,\omega^{2}(t)\,q^2 
\end{equation}
is the standard Lagrangian for the OHDT, where $\omega$ is given by Eq. (\ref{ome}). The addition of a total time-derivative does not modify either the equations of motion or Noether symmetries. 

It is interesting to have the invariants algebra expressing them in terms of $(q, p, t)$ where $p = \dot q - \phi(t)q$ is the canonical momentum and 
\begin{equation}
\{A,B\} = \frac{\partial A}{\partial q}\,\frac{\partial B}{\partial p} - \frac{\partial A}{\partial p}\,\frac{\partial B}{\partial q}
\end{equation}
is the canonical Poisson bracket, where $A = A(q,p,t), B = B(q,p,t)$. The essential results are 
\begin{eqnarray}
\label{ia}
&\strut& 
\{I_1, I_2\} = 1 \,, \quad \{I_1, I_4\} = \{I_3, I_2\} = I_1 \,, \quad  \{I_1, I_5\} = \{I_4, I_2\} = I_2 \,, \\
&\strut& \{I_1, I_3\} = \{I_2, I_5\} = 0 \, \quad \{I_3, I_4\} = I_{1}^2 \,, \quad \{I_3, I_5\} = I_1\,I_2 \,\quad \{I_4, I_5\} = I_2^2 \,. \nonumber 
\end{eqnarray}
Several numerical factors in both Eqs. (\ref{sa}) and (\ref{ia}) are simplified thanks to $g_{1,2}$ having a unit Wronskian. 

\section{Direct approach}

The direct approach towards the explicit symmetry algebra starts from a known gauge field $A(r)$ with the expected asymptotes. Then proceed in a systematic way to the following steps:

\begin{itemize}
\item From $A(r)$, derive $\Phi(r), \phi(t)$ and $\omega(t)$ using resp. Eqs. (\ref{gau}), (\ref{phi}) and (\ref{ome}).
\item Obtain a particular solution $g_1(t)$ solving Eq. (\ref{bb}) and then $g_2$ by the quadrature $(\ref{g2})$.  
\item The Noether symmetry generators are found from Eqs. (\ref{g1})-(\ref{g5}) and invariants from Eqs. (\ref{i1})-(\ref{i5}). 
\end{itemize}





Starting from $A(r)$ or $\phi(t)$ is clearly equivalent. From Eq. (\ref{gau}) one has 
\begin{equation}
n \left(\frac{1}{2} - A(r)\right) = V\,r\,\phi(V\,r) \,,
\end{equation}
where $n$ is an odd integer. From the asymptotes $A(0) = 0 \,, A(\infty) = 1/2$ one should have 
\begin{equation}
t\,\phi(t) \sim \frac{n}{2} \quad {\rm as} \quad t \rightarrow 0 \,,  \quad t\,\phi(t) \rightarrow 0 \quad {\rm as} \quad t \rightarrow \infty \,.
\label{asy}
\end{equation}
%


\section{Example}



From Eq. (\ref{asy}) one generically have $\omega^2 \sim 1/t^2$ as $t \rightarrow 0$, which makes $t = 0$ a singular point of Eq. (\ref{bb}) so that it is difficult to find exact analytical solutions. Nevertheless one can solve  numerically with $g(t_0) = 1, \dot g(t_0) = 0$ (it gives $g_1$) and with $g(t_0) = 0, \dot g(t_0) = 1$ (it gives $g_2$), so that $g_{1,2}$ have unit Wronskian, where $t_0 > 0$ is a reference time. The amplitude of the solution is found to be quite sensitive to the choice of $t_0, n$. Since $\omega^2 \rightarrow 1$ for large $t$ the solution  always becomes sinusoidal with unity angular frequency. On the other hand, it is expressible in terms of Bessel functions for $t \rightarrow 0$ as remarked in \cite{Jackiw}. 

\begin{figure}
    \centering
    \begin{subfigure}[l]{0.45\textwidth}
        \centering
        \includegraphics[width=\linewidth]{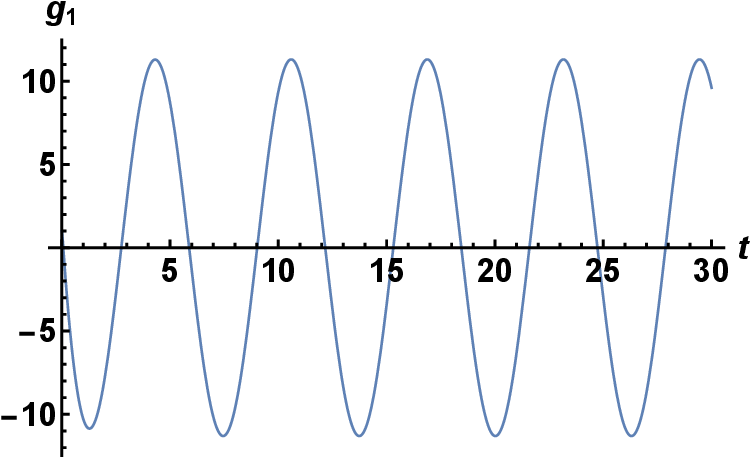} 
    \end{subfigure}
    \begin{subfigure}[r]{0.45\textwidth}
        \centering
        \includegraphics[width=\linewidth]{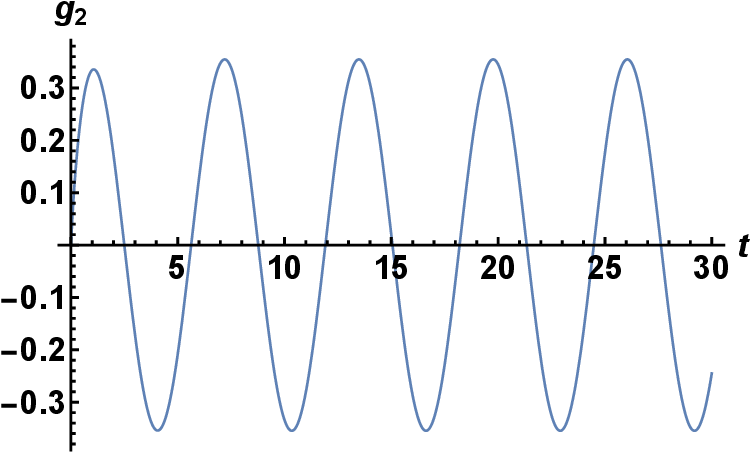} 
    \end{subfigure}
		\caption{Numerical solution of Eq. (\ref{bb}) with $\phi$ given by Eq. (\ref{Bessel}), with $n = 1, t_0 = 10^{-2}$. (a) Left panel: $g(t_0) = 1, \dot g(t_0) = 0$. (b) Right panel: $g(t_0) = 0, \dot g(t_0) = 1$.}
		\label{fig1}
	\end{figure}

\begin{figure}
    \centering
    \begin{subfigure}[l]{0.45\textwidth}
        \centering
        \includegraphics[width=\linewidth]{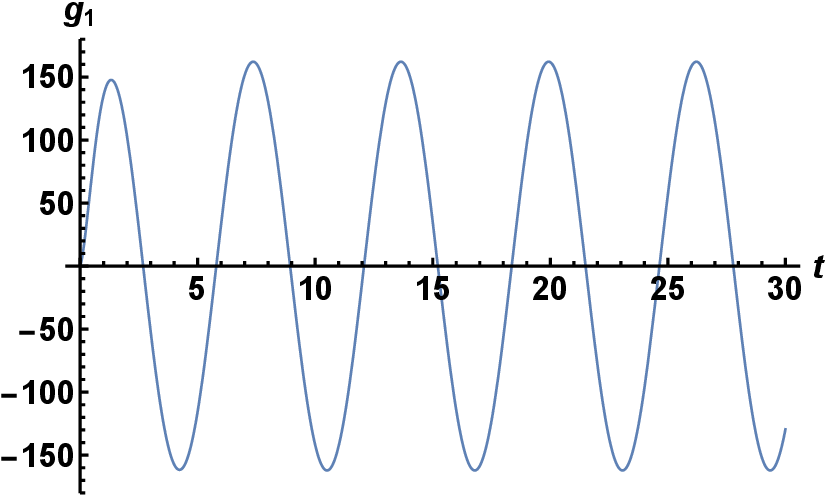} 
    \end{subfigure}
    \begin{subfigure}[r]{0.45\textwidth}
        \centering
        \includegraphics[width=\linewidth]{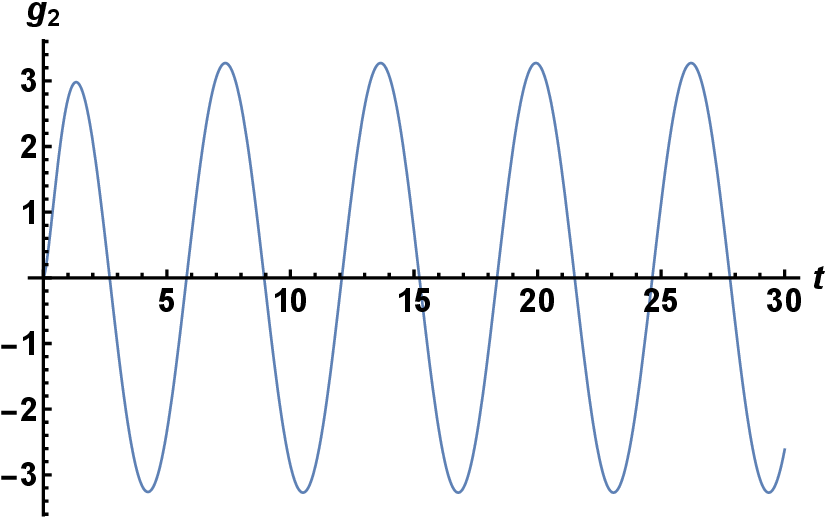} 
    \end{subfigure}
		\caption{Numerical solution of Eq. (\ref{bb}) with $\phi$ given by Eq. (\ref{Bessel}), with $n = 3, t_0 = 10^{-2}$. (a) Left panel: $g(t_0) = 1, \dot g(t_0) = 0$. (b) Right panel: $g(t_0) = 0, \dot g(t_0) = 1$.}
		\label{fig2}
	\end{figure}

\begin{figure}
    \centering
    \begin{subfigure}[l]{0.45\textwidth}
        \centering
        \includegraphics[width=\linewidth]{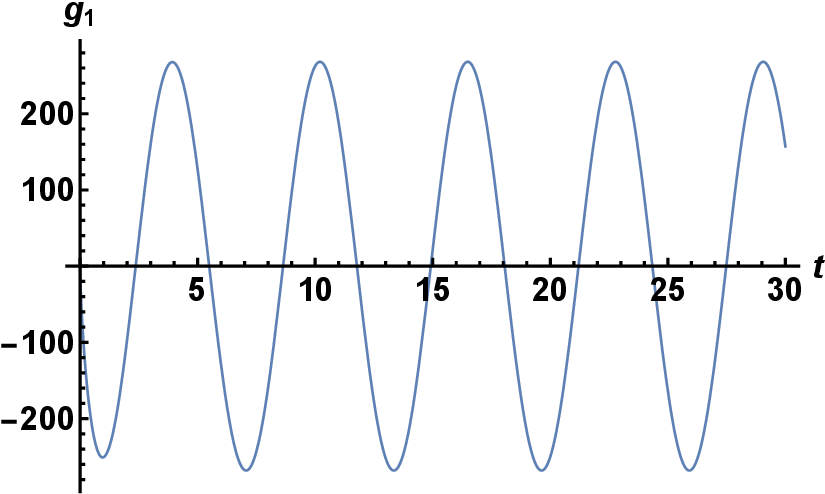} 
    \end{subfigure}
    \begin{subfigure}[r]{0.45\textwidth}
        \centering
        \includegraphics[width=\linewidth]{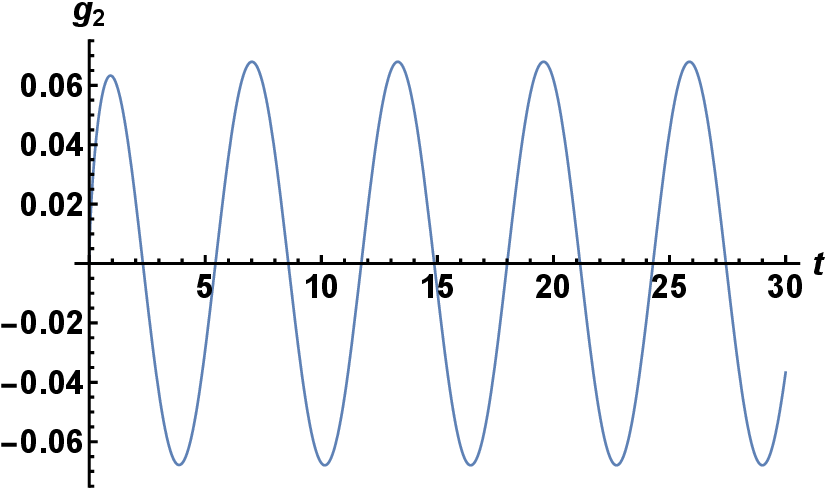} 
    \end{subfigure}
		\caption{Numerical solution of Eq. (\ref{bb}) with $\phi$ given by Eq. (\ref{Bessel}), with $n = 1, t_0 = 10^{-4}$. (a) Left panel: $g(t_0) = 1, \dot g(t_0) = 0$. (b) Right panel: $g(t_0) = 0, \dot g(t_0) = 1$.}
		\label{fig3}
	\end{figure}

For the sake of illustration, we take 
\begin{equation}
\label{Bessel}
\phi = \frac{n}{2}\,K_1(t) \,,
\end{equation}
which satisfies the conditions (\ref{asy}), where $K_1(t)$ is a modified Bessel function of the second kind. This is similar to the profile of vector potentials for decaying magnetic fields in vortex solutions for gauged Ginzburg-Landau theory \cite{Manton}. Examples of numerical solutions are shown in Figs. (\ref{fig1})-(\ref{fig3}) for different $n, t_0$. The details of amplitude and phase can be seen to be very sensitive to $n, t_0$. 



\section{Conclusions}

In this work the Noether point symmetries for the reduced gauged bilayer Jackiw-Pi graphene model have been analyzed. The symmetry algebra is found to be the same as for the TDHO, in view of the existence of an equivalent TDHO Lagrangian. However, in terms of $L$ given by Eq. (\ref{l}) one has $q$ and $p = \dot{q} - \phi q$ directly entering the vortex Dirac spinor as detailed in \cite{Jackiw}, which is not the case if the discussion is made in terms of the standard TDHO Lagrangian ${\cal L}$ in Eq. (\ref{stla}). 

The realization of the symmetry algebra is shown to be achieved in terms of a single particular solution $g_1$ of the linear equation (\ref{bb}), together with $g_2$ which is found from $g_1$ after a quadrature. However, the asymptotes of the gauge field $\phi$ always produce $t = 0$ as a singular point of the TDHO equation (\ref{bb}), so that exact analytical solutions are difficult. The essential third-order linear equation (\ref{aa}) was also shown to be solved from quadratic functions of $g_{1,2}$, as in Eq. (\ref{tt}). In this context this allows to ignore the (nonlinear) Pinney equation, if desired.  The application of the same methods can be a fruitful approach to similar problems such as graphene submitted to external magnetic fields \cite{Kuru}. 








\section*{Acknowledgments} 
The author acknowledges CNPq (Conselho Nacional de Desenvolvimento Cient\'{\i}fico e Tecnol\'ogico) for financial support.


\end{document}